\documentclass[aps,prb,twocolumn,amsmath,amssymb]{revtex4}

\usepackage{times}
\usepackage{graphicx}
\usepackage{amsfonts}
\usepackage{amsmath, amsthm, amssymb}
\usepackage{dsfont}
\usepackage{mathptm}
\usepackage{color}

\newcommand{\eg}{\textit{e.g. }}
\newcommand{\etal}{\emph{et al.}}
\def\i{\mathrm{i}}

\newcommand{\marianne}[1]{\textcolor{blue}{#1}}

\begin{document}

\title{Spin caloritronics with superconductors: \\ enhanced thermoelectric effects, generalized Onsager response-matrix, and thermal spin currents}

\author{Jacob Linder and Marianne Etzelm{\"u}ller Bathen}

\affiliation{Department of Physics, NTNU, Norwegian University of
Science and Technology, N-7491 Trondheim, Norway}

\begin{abstract}
It has recently been proposed and experimentally demonstrated that it is possible to generate large thermoelectric effects in ferromagnet/superconductor structures due to a spin-dependent particle-hole asymmetry. Here, we theoretically show that quasiparticle tunneling between two spin-split superconductors enhances the thermoelectric response manyfold compared to when only one such superconductor is used, generating Seebeck coefficients ($\mathcal{S} > 1$ mV/K) and figures of merit ($ZT \simeq 40$) far exceeding the best bulk thermoelectric materials, and also becomes more resilient toward inelastic scattering processes. We present a generalized Onsager response-matrix which takes into account spin-dependent voltage and temperature gradients. Moreover, we show that thermally induced spin-currents created in such junctions, even in the absence of a polarized tunneling barrier, also become largest in the case where a spin-dependent particle-hole asymmetry exists on both sides of the barrier. We determine how these thermal spin-currents can be tuned both in magnitude and sign by several parameters, including the external field, temperature, and the superconducting phase-difference.
\end{abstract}

\date{\today}

\maketitle

\section{Introduction}
Merging the phenomena of superconductivity and magnetism by creating hybrid structures of materials with these properties is known to give rise to interesting quantum effects \cite{eschrig_physrep_15}. In particular, the field of superconducting spintronics \cite{linder_nphys_15} has in recent years gained increasing attention due to the intriguing prospect of procuring spin-transport with little or no dissipation of energy. In addition to coupling the charge and spin degrees of freedom in such systems, it has in recent developments been shown that adding heat transport to the picture yields surprising new effects \cite{machon_prl_13, kalenkov_prl_12, ozaeta_prl_14, giazotto_apl_14, kalenkov_prb_15, machon_njp_14, kawabata_apl_13, giazotto_prl_15}. A main motivation for the study of the thermoelectricity is that unused waste heat could be utilized as electric currents, and it is desirable to make this conversion process as efficient as possible.

It was theoretically proposed in Ref. \cite{machon_prl_13} that by lifting the spin degeneracy of the density of states in superconductors (\eg by proximity to magnetic materials), very large thermoelectric effects could be achieved. Ref. \cite{kalenkov_prl_12} showed that an electron-hole asymmetry induced by magnetic impurities in superconductors could lead to sizable thermoelectric currents. Following works demonstrated how it was possible to achieve even higher thermoelectric figures of merit $ZT$ and Seebeck coefficients $\mathcal{S}$ by making use of the large accumulation of quasiparticle states at energies near the gap edge ($E \simeq \Delta_0$) in superconductors \cite{ozaeta_prl_14, giazotto_apl_14}. Large thermophases induced in magnetic Josephson junctions have also been studied \cite{giazotto_prl_15}. The usage of superconducting elements in low-temperature thermometry and refrigeration has been studied extensively in the past \cite{giazotto_rmp_06}, but it is only quite recently that the incorporation of magnetic elements into such structures has sparked considerable interest.

The strong coupling of spin, heat, and charge transport in superconducting structures allows us to envision a number of interesting cryogenic thermoelectric devices exceeding the performance of their non-superconducting counterparts, such as highly sensitive thermal sensors. A recent preprint \cite{kolenda_arxiv_15} reported experimental observation of the large thermoelectric currents predicted in \cite{ozaeta_prl_14} by utilizing an normal metal/ferromagnetic barrier/superconductor junction (Cu/Fe/Al). Upon application of strong in-plane magnetic fields $B \sim 1$ T, Seebeck coefficients $|\mathcal{S}|$ up to 0.1 mV/K were measured. The key to achieving this effect is to create a spin-dependent particle-hole asymmetry in the superconductor (Al) by applying an in-plane field. Due to the magnetic barrier (Fe), tunneling of one spin species is favored compared to the other, thus effectively probing the energy asymmetry for each spin $\sigma$. This scenario raises a tantalizing question: what happens if a spin-dependent particle-hole asymmetry exists not only on one side of the magnetic barrier, but on both sides? One might expect that creating such an asymmetry in all regions of the system would strongly enhance thermoelectric effects even beyond what has been predicted so far for bilayer structures.

\begin{figure*}[t!]
\begin{centering}
\includegraphics[width=0.5\textwidth]{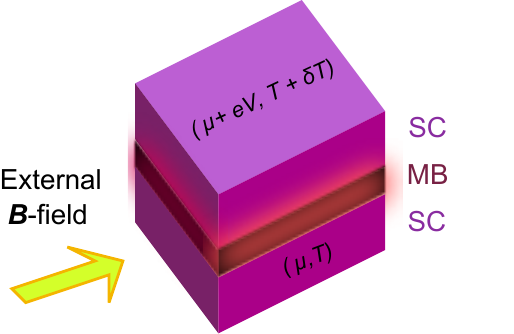}\includegraphics[width=0.47\textwidth]{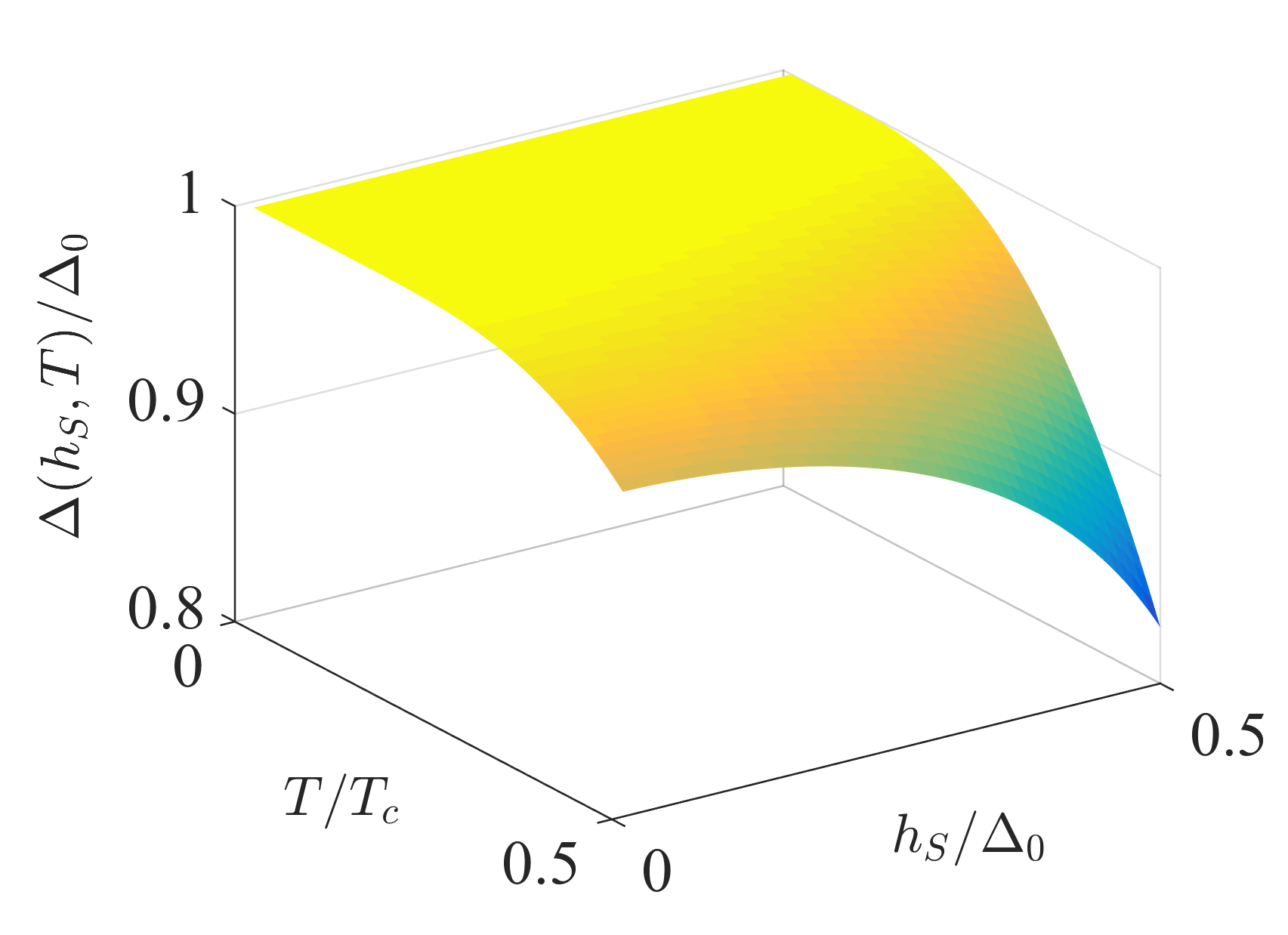}
\caption{(Color online) \textit{Left panel}: quasiparticle tunneling between 
two thin superconductors (SC) $S_L$ and $S_R$ separated by a magnetic barrier (MB). 
A spin-dependent particle-hole asymmetry is induced 
via an external in-plane magnetic field. \textit{Right panel}: 
Self-consistent solution of the order parameter in a 
spin-split superconductor as a function of exchange field $h_S$ and temperature $T$.}
\label{fig:model}
\end{centering}
\end{figure*}

In this work, we confirm this hypothesis and show that quasiparticle tunneling between two spin-split superconductors not only increases the thermoelectric response of the system manyfold, but importantly also displays a robustness toward inelastic scattering in the system. The latter aspect is of particular importance with regard to material choice and possible use of thermoelectric effects in cryogenic devices. For instance, Al is known to have a weak inelastic scattering rate (modelled by \eg a Dynes \cite{dynes} parameter $\Gamma$), but also has a very low critical temperature $T_c = 1.2$ K. By achieving large thermoelectric effects even at considerable inelastic scattering $\Gamma$, it becomes possible to use superconductors with much higher critical temperatures such as NbN featuring $T_c=14$ K. Our results therefore provide a way in which robust spin caloritronics with superconductors can be achieved above the sub-Kelvin regime, featuring figures of merit up to $ZT\simeq 40$ and Seebeck coefficients $\mathcal{S} > 1$ mV/K which far exceeds even the best thermoelectric bulk materials such as CsBi$_4$Te$_8$ and Bi$_2$Te$_3$ that have $ZT \simeq 2$ at room-temperature \cite{snyder_natmat_08}.

Previous works \cite{machon_prl_13, ozaeta_prl_14} have considered how voltage and temperature gradients induce thermoelectric effects in superconducting junctions where spin-degeneracy is lifted. Here, we present a generalized Onsager response-matrix which takes into account the possibility of having spin-dependent voltages and temperature-biases. The latter scenarios can be realized through tunneling between ferromagnetic and non-magnetic materials, as predicted in Refs. \cite{hatami_prl_07, veramarun_prl_14, heikkila_prb_10}.  In Ref. \cite{dejene_nature_16} a spin-dependent heat conductance was observed in F/N/F spin-valve nanopillars. This was assumed to arise due to spin heat accumulation, and a difference in effective spin temperature of up to 350 mK was reported. The same effect was also observed in Ref. \cite{kimling_prb_15} more recently, but the authors were in this case more reluctant in concluding that the observation did in fact prove the existence of spin heat accumulation.

Hybrid structures with spin-split superconductors admit thermally induced spin-currents, without requiring any polarized barrier as noted in Ref. \cite{ozaeta_prl_14}, yet this phenomenon has not yet been studied in detail. We demonstrate that these spin-currents are in fact the largest precisely in the case where a spin-dependent particle-hole asymmetry exists on both sides of the barrier. Moreover, we determine how these thermal spin-currents can be tuned, both in magnitude and sign, by several parameters, including the external field, temperature, and the superconducting phase-difference when incorporating Josephson junctions into the geometry.

\begin{figure*}[t!]
\includegraphics[width=1.01\textwidth]{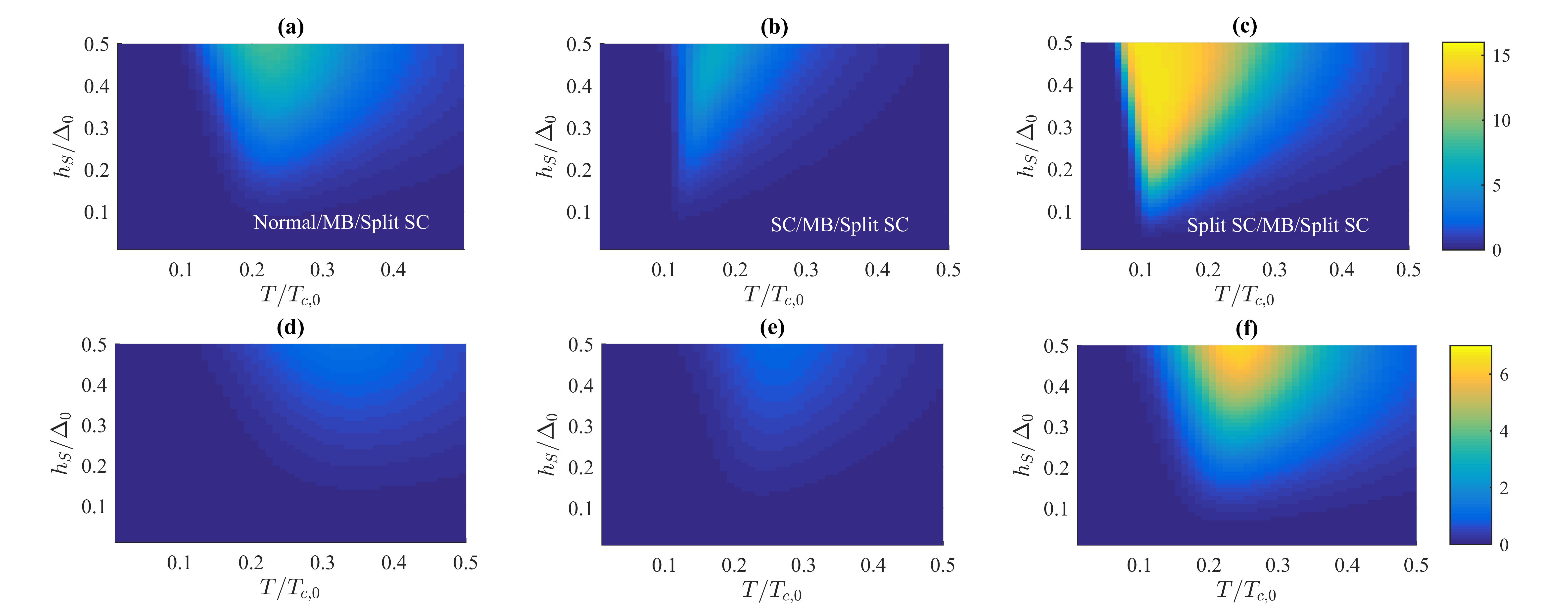}
\caption{(Color online) Figure of merit $ZT$ in the $h_S$-$T$ plane for a barrier polarization $P=97\%$. In the top row, $\Gamma/\Delta_0=10^{-3}$ while in the bottom row the inelastic scattering is substantial $\Gamma/\Delta_0=0.05$. Three bilayer setups separated by a magnetic barrier (MB) are compared: \textbf{(a) \& (d)} normal metal / MB / spin-split superconductor, \textbf{(b) \& (e)} superconductor / MB / spin-split superconductor,  \textbf{(c) \& (f)} spin-split superconductor / MB / spin-split superconductor. As seen, the thermoelectric response is dramatically enhanced in the last case.}
\label{fig:results}
\end{figure*}

 \section{Theory}\label{sec:theory}
The system under consideration is shown in Fig. \ref{fig:model}(a) and consists of two spin-split superconductors separated by a magnetic barrier with an in-plane magnetic field applied. Possible material choices could be Al/Fe/Al, along the lines of Ref. \cite{kolenda_arxiv_15}, but NbN/GdN/NbN might be more beneficial due to the strong polarization of GdN \cite{pal_natcom_14} and high $T_c$ of NbN. An additional advantage of using a more strongly polarized ferromagnetic barrier is that it can by itself induce an exchange field into both of the superconductors \cite{santos_prl_08}, necessitating lower externally applied fields. If desirable, one can substitute one of the superconducting electrodes with a thin normal metal in proximity to a superconducting film, in which case the normal metal mimicks a spin-split superconductor in the presence of an in-plane field $\boldsymbol{B}$. When Coulomb blockade and the supercurrent response is suppressed, quasiparticle tunneling dominates the transport across the junction \cite{meservey}. We here seek to establish a spin-dependent particle-hole asymmetry throughout the system which is accomplished by using not just a single spin-split superconductor, as in \eg\cite{ozaeta_prl_14, giazotto_apl_14, kolenda_arxiv_15}, but two. In this way, both electrodes $S_L$ and $S_R$ outlined in  Fig. \ref{fig:model}(a) host a large particle-hole asymmetry for spin $\sigma$. Because of this, a crucial effect comes into play: since now the asymmetry exists on both sides of the junction, an additional term appears in the thermoelectric currents as we will show below. We will also demonstrate that large thermoelectric effects are retained in the proposed setup even in the presence of substantial inelastic scattering.

An important point which should be emphasized is the role of phonon 
contribution to the thermal conductance, which is known to be 
important for semiconducting thermoelectric materials. In 
contrast, in metals the heat transfer by electrons strongly 
dominates over the phonon contribution at low temperatures. However, 
in the superconducting state the electron contribution  
decreases with temperature due to the exponential decrease in the carrier density, 
while the phonon contribution increases due to suppression of the 
phonon-electron scattering. Therefore, a model neglecting phonon 
heat transfer in the superconducting bulk becomes less applicable as $T\to 0$. For specific superconducting materials, the model 
applicability requires a detailed comparison of electron and phonon 
thermal conductivities.

The charge and heat tunneling currents carried by spin-species $\sigma$ read \cite{mahan_book, heikkila_book}:
\begin{align}\label{eq:currents}
I_\text{heat}^\sigma &= \frac{G_\sigma}{e^2} \int^\infty_{-\infty} dE (E-\mu_L)\mathcal{D}_L^\sigma(E-\mu_L) \mathcal{D}_R^\sigma(E) F(E),\notag\\
I_\text{charge}^\sigma &= \frac{G_\sigma}{e} \int^\infty_{-\infty} dE \mathcal{D}_L^\sigma(E-\mu_L) \mathcal{D}_R^\sigma(E) F(E),
\end{align}
where $I_\text{heat}^\sigma$ is the heat current flowing out of the left electrode. Here, the quasiparticle energy $E$ is measured relative the Fermi level in the right superconductor, $\mu_L$ is the Fermi level in the left region ($\mu_R=0$ for reference), $\mathcal{D}_j^\sigma$ is the density of states for spin $\sigma$ in region $j$, $f_j(E)$ is the distribution function in region $j$, and $F=f_L(E-\mu_L)-f_R(E)$. The superconducting regions are assumed to have a small thickness ($t \sim 10-20$ nm) as in the experiment of Ref. \cite{kolenda_arxiv_15}, so that an externally applied field splits the density of states according to 
\begin{align}
\mathcal{D}^\sigma = \Bigg|\text{Re}\Bigg\{  \frac{E + \sigma h_S + \i\Gamma}{\sqrt{(E+\sigma h_S+\i\Gamma)^2 - \Delta^2}}  \Bigg\}\Bigg|
\end{align}
with $h_S$ being the induced Zeeman-field in S and $\Delta=\Delta(h_S,T)$ is the superconducting gap. Its dependence on $h_S$ and $T$ is shown in Fig. \ref{fig:model}(b), featuring a first order phase-transition at $(h/\Delta_0, T/T_{c,0}) = (0.52,0.53)$ where $\Delta_0$ and $T_{c,0}$ is the bulk superconducting gap and critical temperature in the absence of the field. Interfacial spin-flip scattering would be likely to reduce the net barrier polarization effect due to the randomization of spin.

\begin{figure*}[t!]
    \includegraphics[width=0.45\textwidth]{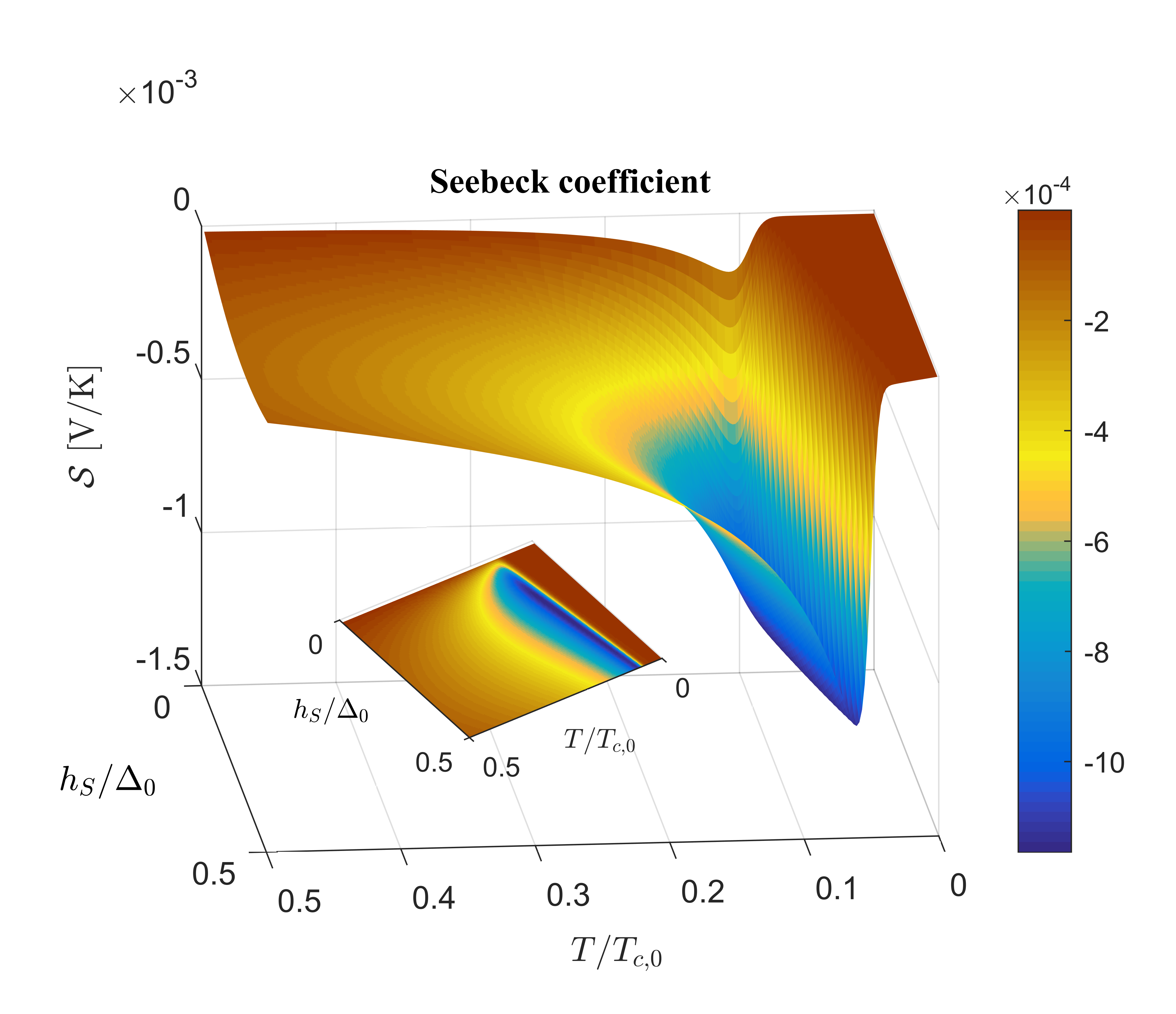}\includegraphics[width=0.55\textwidth]{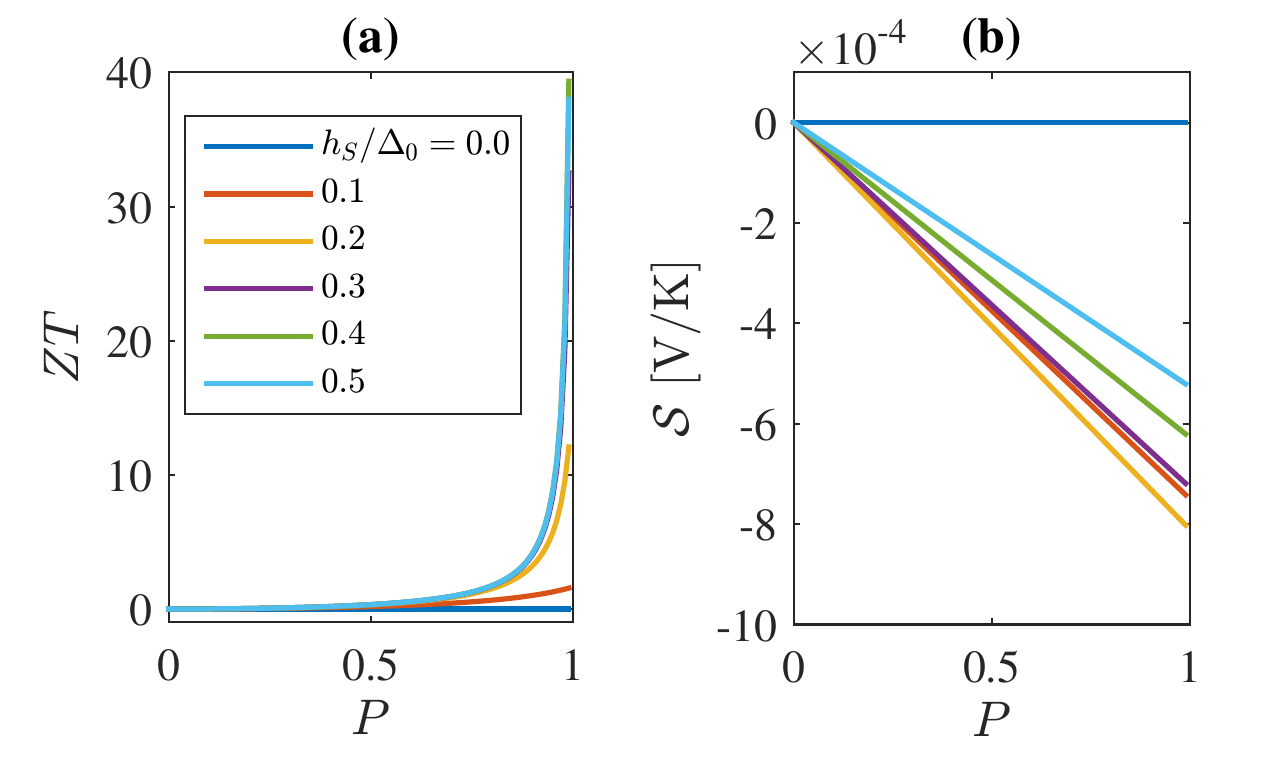}
    \caption{(Color online) \textit{Left panel:} Seebeck coefficient $\mathcal{S}$ for quasiparticle 
tunneling between two spin-split superconductors separated 
by a magnetic barrier. We set $P=97\%$ and $\Gamma/\Delta_0 = 10^{-3}$. The inset shows a bird's eye view of the same plot. \textit{Right panels:} \textbf{(a)} The figure of merit and \textbf{(b}) Seebeck coefficient for the same setup as a function of the barrier polarization $P$. We have set $T/T_{c,0}=0.15$ and $\Gamma/\Delta_0=10^{-3}$. Close to $P=1$, figures of merit $ZT\simeq 40$ are obtained.}
    \label{fig:seebeck}
\end{figure*}

\section{Results}\label{sec:results}
\subsection{Thermoelectric figure of merit and Seebeck coefficient}
In the presence of a voltage difference $V$ or temperature gradient $\Delta T$ across the bilayer, the Onsager matrix equation \cite{onsager} describing the linear response for the total charge $I=I_\text{charge}^\uparrow+I_\text{charge}^\downarrow$ and heat current $\dot{Q} = I_\text{heat}^\uparrow + I_\text{heat}^\downarrow$ flowing through the interface reads:
\begin{align}
\begin{pmatrix}
I \\
\dot{Q} \\
\end{pmatrix} = 
\begin{pmatrix}
L_{11} & L_{12} \\
L_{12} & L_{22} \\
\end{pmatrix}\begin{pmatrix}
V \\
\Delta T /T\\
\end{pmatrix},
\end{align}
where we have used that $L_{21}=L_{12}$ due to symmetry (as can also be proven analytically). We defined here $\Delta T/2T = (T_L-T_R)/(T_L+T_R)$. To identify the Onsager coefficients $L_{ij}$, one performs an expansion of Eq. (\ref{eq:currents}) to lowest order in applied voltage $V$ and temperature gradient $\Delta T$, which after some algebra yields the result:
\begin{align}\label{eq:onsager}
L_{11} &= G_T \int^\infty_{-\infty} dE(\mathcal{D}_L^0 \mathcal{D}_R^0 + \mathcal{D}_L^z\mathcal{D}_R^z/4)C(E),\notag\\
L_{22} &= \frac{G_T}{e^2}\int^\infty_{-\infty} dE(\mathcal{D}_L^0 \mathcal{D}_R^0 + \mathcal{D}_L^z\mathcal{D}_R^z/4)E^2C(E),\notag\\
L_{12} &= \frac{G_TP}{2e} \int^\infty_{-\infty} dE(\mathcal{D}_L^0 \mathcal{D}_R^z + \mathcal{D}_L^z\mathcal{D}_R^0)EC(E),
\end{align}
with $C(E) = [4 k_BT\cosh^2(\beta E/2)]^{-1}$. We have here defined:
\begin{align}
\mathcal{D}^0_j &= (\mathcal{D}_j^\uparrow + \mathcal{D}_j^\downarrow)/2,\notag\\
\mathcal{D}^z_j &= \mathcal{D}_j^\uparrow - \mathcal{D}_j^\downarrow
\end{align}
for side $j\in\{L,R\}$ In previous proposals, a spin-dependent particle-hole asymmetry existed only in $S_R$ while a metal \cite{ozaeta_prl_14} or a superconductor with a tunable gap \cite{giazotto_apl_14} was used in the place of $S_L$. However, in the present case the asymmetry of the structure is maximized in the sense that it exists on both sides of the interface and, importantly, generates additional terms in the Onsager coefficients as shown in Eq. (\ref{eq:onsager}). For instance, the coefficient $L_{12}$ responsible for inducing heat flow due to a voltage gradient (and also an electric current due to a temperature gradient) now couples the antisymmetric (in $E$) component $\mathcal{D}^z$ on the left side of the magnetic barrier to the symmetric component $\mathcal{D}^0$ of the right side and vice versa. This strongly modifies the thermoelectric response of the system. Of particular interest are the Seebeck coefficient $\mathcal{S}$ (the voltage induced due to a temperature difference after opening the circuit) and the dimensionless figure of merit $ZT$ (which quantities the ability of the system to produce thermoelectric power efficiently) \cite{1964}:
\begin{align}
\mathcal{S} = -\frac{L_{12}}{L_{11} T},\; ZT = \Big(\frac{L_{11}L_{22}}{L_{12}^2} - 1\Big)^{-1}.
\end{align}
We now proceed to show that due to the additional spin-splitting in $S_L$ ($\mathcal{D}_L^z\neq0)$, the thermoelectric effects are enhanced manyfold compared to when a metal or conventional superconductor is used and that they remain large even in the presence of substantial inelastic scattering $\Gamma$.

\begin{figure}[b!]
    \includegraphics[width=0.5\textwidth]{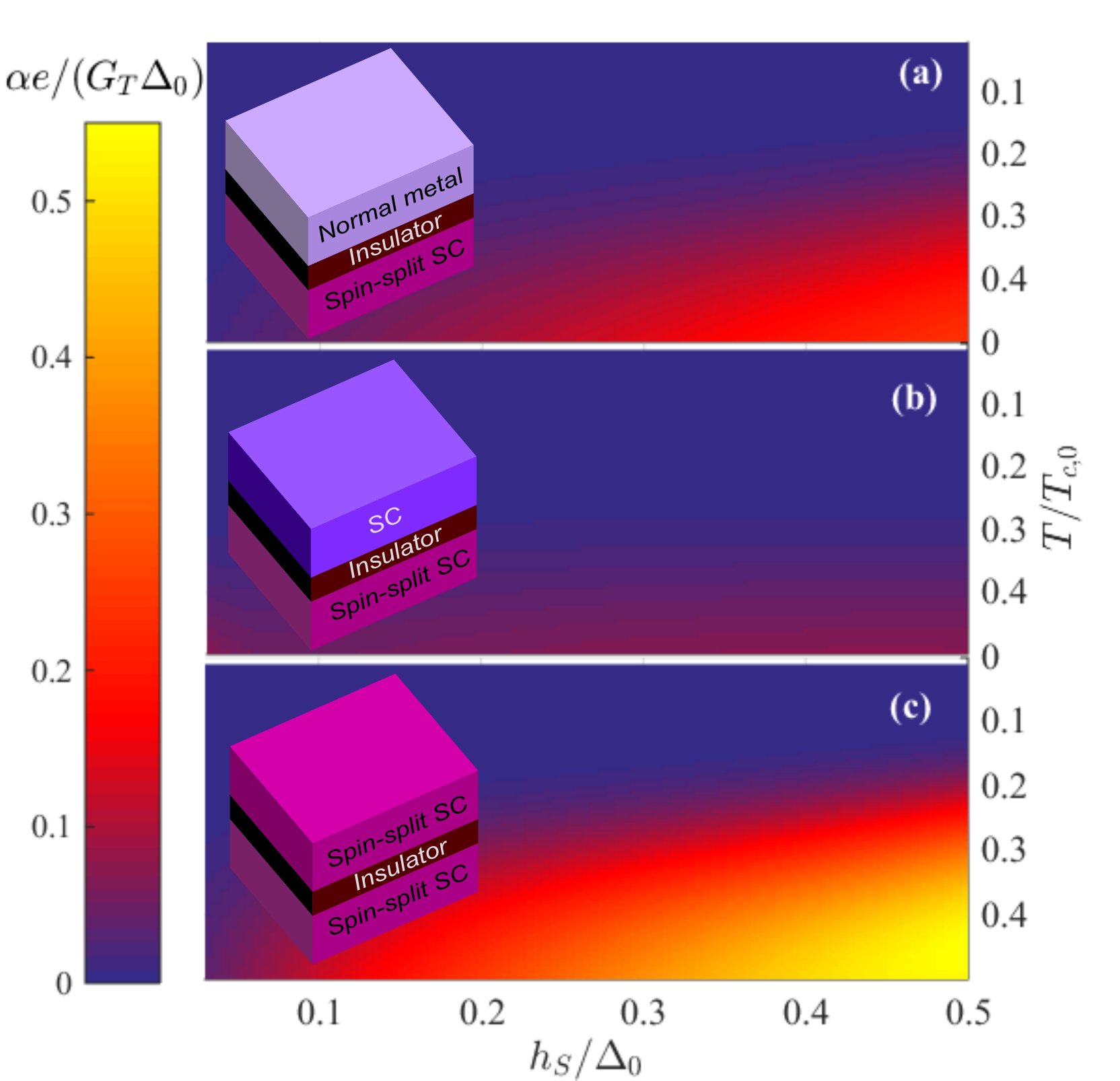}
    \caption{(Color online) Plot of the normalized thermoelectric response coefficient $(\alpha e)/(G_T\Delta_0)$ that governs the thermally induced spin current for bilayer junctions without any polarizing barrier: (a) normal metal/insulator/spin-split SC, (b) SC/insulator/spin-split SC, and (c) spin-split SC/insulator/spin-split SC.}
    \label{fig:alpha_bilayer}
\end{figure}

In Fig. \ref{fig:results}(a)-(c), we have plotted the thermoelectric figure of merit $ZT$ obtained as a function of temperature $T/T_{c,0}$ and exchange field $h_S/\Delta_0$ upon using a magnetic barrier with polarization $P=0.97$ (as suitable for \eg GdN \cite{pal_prb_15}) and with inelastic scattering $\Gamma/\Delta_0=10^{-3}$. Extraordinarly large figures of merit $ZT>15$ are obtained when the quasiparticle tunneling occurs between two spin-split superconductors as shown in Fig. \ref{fig:results}(c). In comparison, the best thermoelectric materials at room-temperature (CsBi$_4$Te$_8$ and Bi$_2$Te$_3$) reach $ZT \simeq 2$. When only one spin-split superconductor is used \cite{ozaeta_prl_14, giazotto_apl_14}, the thermoelectric response is much smaller as seen in (a) and (b). For smaller polarization values $P$, the figure of merit $ZT$ is suppressed for every type of hybrid structure but still remains largest for tunneling between two spin-split superconductors. The precise dependence on the barrier polarization is shown in Fig. \ref{fig:seebeck}(a). As $P$ increases, $ZT$ becomes colossal and reaches almost 40 in magnitude. Since the exchange splitting of the density of states in the superconductors is tunable via an external field, it should be possible to exert well-defined control over the thermoelectric response of the system.

\begin{figure*}[t!]
    \includegraphics[width=0.99\textwidth]{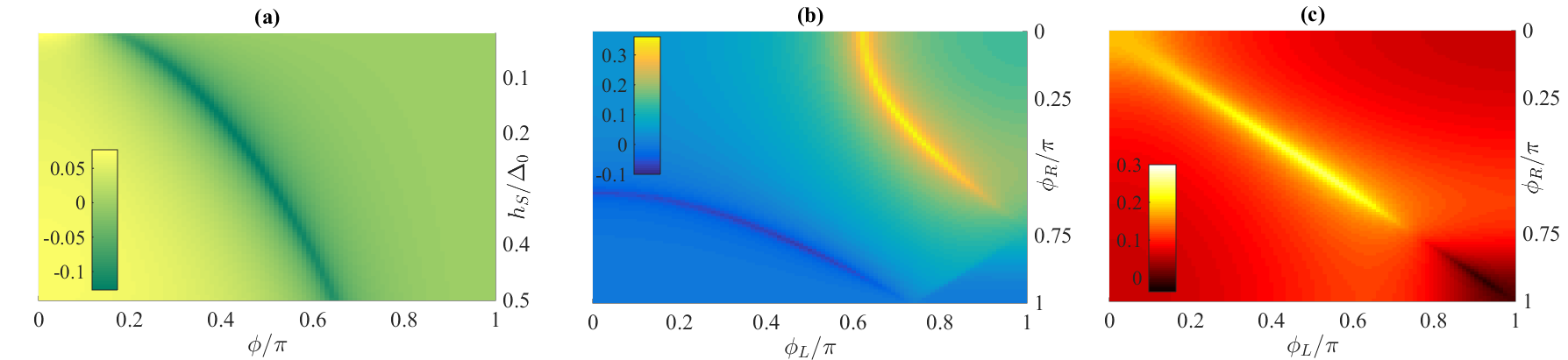}
    \caption{(Color online) Plot of the normalized thermoelectric response coefficient $(\alpha e)/(G_T\Delta_0)$ that governs the thermally induced spin current for three types of structures. (a) Tunneling between a superconducting electrode and the normal part of a spin-split SC/normal/spin-split SC junction. We have set $T/T_{c,0}=0.5$.  The sign of $\alpha$ can be changed, inverting the direction of the spin current flow, by tuning $h_S$ or the superconducting phase difference $\phi$. (b) Tunneling between the normal parts of a SC/N/SC and a spin-split SC/N/spin-split SC junction, with $T/T_{c,0}=0.4$ and $h_S/\Delta_0=0.4$. Two arcs with opposite signs cross the $\phi_L$-$\phi_R$ parameter space, where $\phi_L$ is the phase difference between the SCs and $\phi_R$ is the phase difference between the spin-split SCs. (c) Tunneling between the normal parts of two spin-split SC/N/spin-split SC junctions, having set $T/T_{c,0}=0.4$ and $h_S/\Delta_0=0.2$. The normal layers are all assumed to be short compared to the penetration depth of the superconducting correlations, so that they become fully proximitized. We acknowledge the challenge in experimentally realizing tunneling between the weak links of two Josephson junctions, as in (b) and (c), but we nevertheless include these results to demonstrate the interesting behavior of the thermal spin current in this scenario.  }
    \label{fig:total}
\end{figure*}

In order to demonstrate the robustness of the results toward inelastic scattering, we plot in Fig. \ref{fig:results}(d)-(f) the figure of merit $ZT$ for a 50 times larger inelastic scattering rate $\Gamma/\Delta_0=0.05$. This amounts to quite heavy suppression of the BCS coherence peaks in the density of states and smooths out spectral features greatly. In spite of this, it is seen that in the case of quasiparticle tunneling between two spin-split superconductors a figure of merit close to $ZT>5$ is retained [Fig. \ref{fig:results}(f)] whereas in the other cases $ZT$ is close to an order of magnitude smaller. Another measure of the efficiency of thermoelectric effects is the Seebeck coefficient $\mathcal{S}$, and we plot its behavior in Fig. \ref{fig:seebeck} for the setup shown in Fig. \ref{fig:model}(a). Magnitudes of $|\mathcal{S}| >  1$ mV/K are attainable, which is an order of magnitude larger than in the experiment of Ref. \cite{kolenda_arxiv_15} where only one spin-split superconductor was used. It should be noted that a rather weak polarization $P \simeq 0.1$ was utilized in Ref. \cite{kolenda_arxiv_15} and for larger polarizations $\mathcal{S}$ could theoretically reach the order of 1 mV/K in such a setup as well by fine-tuning the parameters.

\subsection{Generalized Onsager response-matrix}
Besides applying a voltage or temperature bias, it is also experimentally feasible to create a spin-dependent voltage and temperature bias, $V_s$ and $\Delta T_s$, respectively.  Tunneling between ferromagnetic materials and non-magnetic conductors has been predicted to result in spin-dependent effective temperatures and voltages, and recent experimental results support these claims \cite{dejene_nature_16}. This would allow for the application of spin-dependent biases through the addition of ferromagnetic layers to one of the electrodes, and heating these to different temperatures.  In the presence of spin-dependent gradients, the Onsager response-matrix is generalized to
\begin{align}\label{eq:general}
\begin{pmatrix}
I \\
\dot{Q} \\
I_s \\
\dot{Q}_s \\
\end{pmatrix} = 
\begin{pmatrix}
G & P\alpha & PG & \alpha \\
P\alpha & G_Q & \alpha & PG_Q \\
PG & \alpha & G & P\alpha \\
\alpha & PG_Q & P\alpha & G_Q\\
\end{pmatrix}\begin{pmatrix}
V \\
\Delta T /T\\
V_s/2\\
\Delta T_s /(2T)\\
\end{pmatrix} \marianne{.}
\end{align}
Above, we have defined the spin current $I_s=I_\text{charge}^\uparrow-I_\text{charge}^\downarrow$ and spin heat current $\dot{Q}_s = I_\text{heat}^\uparrow - I_\text{heat}^\downarrow$. The applied voltage and temperature biases are in the spin-dependent case given by:
\begin{align}
V &= \sum_\sigma (V_L^\sigma - V_R^\sigma)/2,\; V_s = \sum_\sigma \sigma(V_L^\sigma - V_R^\sigma),\notag\\
\Delta T &= \sum_\sigma (T_L^\sigma - T_R^\sigma)/2,\; \Delta T_s = \sum_\sigma \sigma(T_L^\sigma - T_R^\sigma),
\end{align}
and $T = \sum_\sigma (T_R^\sigma + T_L^\sigma)/4$. In order to simplify the expressions, spin-dependent biases were assumed to exist only on the left-hand side of the barrier. Consequently, we have defined $T_R^\uparrow = T_R^\downarrow = T_R$ and $V_R^\uparrow = V_R^\downarrow = 0$ for reference. The thermoelectric coefficients in Eq. (\ref{eq:general}) read $G = L_{11}, G_Q = L_{22},$ and
\begin{align}\label{eq:alpha}
\alpha = \frac{G_T}{2e} \int^\infty_{-\infty} dE(\mathcal{D}_L^0 \mathcal{D}_R^z + \mathcal{D}_L^z\mathcal{D}_R^0)EC(E).
\end{align}
This reveals some interesting cross-couplings between spin and heat flow that exist due to the spin-dependent particle-hole asymmetry induced in the superconductors by an exchange field. For instance, one can obtain a heat current $\dot{Q}$ by applying a spin-dependent voltage $V_s$. The response-matrix presented above is general, as it allows for arbitrary voltage and temperature differences for each spin.

\subsection{Thermally induced spin-currents}
Equation (\ref{eq:general}) shows that even in the absence of any barrier polarization in the junction $(P=0)$, a spin-current $I_s$ can be induced via a temperature-gradient $\Delta T$ without any accompanying charge flow, according to $I_s = \alpha\Delta T/T$. This fact was also noted in Ref. \cite{ozaeta_prl_14}. We emphasize that this thermal spin-current will also flow in the bulk of the superconductor since it is carried by spin-polarized quasiparticles. Up to now, this phenomenon has not been studied in detail and we therefore determine in what follows how this spin current can be controlled both in magnitude and in sign by using hybrid structures with spin-split superconductors. The quantity of interest is thus the thermoelectric coefficient $\alpha$ in Eq. (\ref{eq:alpha}) and in what follows we compute it numerically for several types of hybrid structures, setting $\Gamma/\Delta_0=0.005$. 

We start by comparing in Fig. \ref{fig:alpha_bilayer} the thermal spin-current for the same structures as in Fig. \ref{fig:results} (normal/spin-split SC, SC/spin-split SC, and spin-split SC/spin-split SC), but now with the absence of any polarizing barrier $(P=0)$. The resulting $\alpha$ is by far the largest in case (c), demonstrating again the advantage in creating a spin-dependent particle-hole asymmetry on both sides of the interface. By incorporating a Josephson junction in the geometry, the superconducting phase difference becomes an additional external control parameter that can be used to adjust the thermal spin current, similarly to the setup of Ref. \cite{giazotto_apl_14}. We find that not only the magnitude of $\alpha$, and in turn $I_s$, but also its sign can be changed. This is shown in Fig. \ref{fig:total}, where we plot the normalized thermoelectric coefficient $(\alpha e)/(G_T\Delta_0)$ for various types of hybrid structures incorporating spin-split superconductors. Varying the precise values of $h_S$ and $T$ produce qualitatively similar plots in all cases, and thus we show only one representative plot for each type of system in Fig. \ref{fig:total}. The thermal spin current responds to a change in the superconducting phase difference $\phi$ since the proximity-induced minigap $\Delta_g$ in the normal metal region depends on it via $\Delta_g = \Delta(h,T)\cos(\phi/2)$, where $\Delta(h,T)$ is the gap in the bulk superconductors of the Josephson contact. Fig. \ref{fig:total} demonstrates that the thermal spin current demonstrates a rich variety of qualitative behavior, depending on the type of structure that is used.

\section{Concluding remarks}
The above results thus show that spin-dependent thermoelectric effects in superconductors are increased when a spin-dependent particle-hole asymmetry exists in both adjacent layers to a magnetic tunneling barrier. Coupling spin and heat transport is the foundation for spin caloritronics \cite{bauer_natmat_12}, which suggests that highly sensitive thermoelectric elements can be tailored by using superconductors, leading to efficiencies far exceeding what is possible in non-superconducting materials. An interesting future direction could be to explore the role of unconventional superconducting pairing symmetries combined with magnetic elements with regard to thermoelectric effects \cite{lofwander_prb_04}, such as the $d$-wave pairing of the high-$T_c$ cuprates or $p$-wave pairing in the uranium-based ferromagnetic superconductors. The study of Josephson junction geometries is also of interest: by combining such a setup with one spin-split superconductor so that a proximity-induced superconducting gap can be tuned, the figure of merit can under ideal circumstances become comparable \cite{giazotto_apl_14} to the present case with tunneling between two spin-split superconductors. Moreover, the existence of strong odd-frequency triplet pairing in spin-split superconductors was recently highlighted \cite{linder_scirep_15} and suggests that other systems where odd-frequency superconducting pairing is present and renders the electronic density of states spin-dependent, such as junctions with magnetic spin-valves \cite{halterman_prl_07, karminskaya_prb_11}, spin-active interfaces \cite{machon_njp_14, asano_prb_07, gomperud_prb_15, triola_prb_14, terrade_prb_13}, inhomogeneous magnetization \cite{cottet_prl_11, dibernardo_natcom_15}, or spin-orbit coupling \cite{jacobsen_prb_15a, jacobsen_prb_15b, arjoranta_prb_16, gentile_prl_13}, could host large thermoelectric effects as well. We leave these prospects for forthcoming studies.\\
\textbf{ }\\

\acknowledgements

 J.L was supported by the Research Council of Norway, Grants No. 205591,
216700, 240806 and the ”Outstanding Academic Fellows” programme at NTNU. T. Heikkil{\"a} is thanked for useful comments.

\end{document}